\documentclass[10pt]{iopart}

\expandafter\let\csname equation*\endcsname=\relax
\expandafter\let\csname endequation*\endcsname=\relax
\usepackage{amsmath}
\usepackage{iopams}  
\usepackage{float}
\usepackage{txfonts}
\usepackage{array}
\newcolumntype{C}[1]{>{\centering\arraybackslash}p{#1}}

\usepackage{graphicx}

\usepackage{subcaption}  
\usepackage{pdfpages}
\usepackage{caption}
\usepackage{cite}

\DeclareCaptionFormat{myformat}{\textbf{#1}#2#3}
\begin{document}
\frenchspacing

\title[Adapted Swin Transformer-based Real-Time Plasma Shape Detection and Control in HL-3]{Adapted Swin Transformer-based Real-Time Plasma Shape Detection and Control in HL-3}



\author{Qianyun Dong$^{1}$, Zhengwei Chen$^{2}$, Rongpeng Li$^{1,*}$, Zongyu Yang$^{2}$, Feng Gao$^{3}$, Yihang Chen$^{2}$, Fan Xia$^{2}$, Wulyu Zhong$^{2,*}$, Zhifeng Zhao$^{3}$}
\address{1. Zhejiang University, Hangzhou 310058, China}
\address{2. Southwestern Institute of Physics, Chengdu 610043, China}
\address{3. Zhejiang Lab, Hangzhou 311500, China}
\ead{lirongpeng@zju.edu.cn, zhongwl@swip.ac.cn}

\begin{abstract}

In the field of magnetic confinement plasma control, the accurate feedback of plasma position and shape primarily relies on calculations derived from magnetic measurements through equilibrium reconstruction or matrix mapping method. However, under harsh conditions like high-energy neutron radiation and elevated temperatures, the installation of magnetic probes within the device becomes challenging. Relying solely on external magnetic probes can compromise the precision of EFIT in determining the plasma shape. To tackle this issue, we introduce a real-time, non-magnetic measurement method on the HL-3 tokamak, which diagnoses the plasma position and shape via imaging. Particularly, we put forward an adapted Swin Transformer model, the Poolformer Swin Transformer (PST), to accurately and fastly interpret the plasma shape from the Charge-Coupled Device Camera (CCD) images. By adopting multi-task learning and knowledge distillation techniques, the model is capable of robustly detecting six shape parameters under disruptive conditions such as a divertor shape and gas injection, circumventing global brightness changes and cumbersome manual labeling. Specifically, the well-trained PST model capably infers $R$ and $Z$ within the mean average error below $1.1$ cm and $1.8$ cm, respectively, while requiring less than $2$ ms for end-to-end feedback, an $80\%$ improvement over the smallest Swin Transformer model, laying the foundation for real-time control. Finally, we deploy the PST model in the Plasma Control System (PCS) using TensorRT, and achieve $500$ ms stable PID feedback control based on the PST-computed horizontal displacement information. In conclusion, this research opens up new avenues for the practical application of image-computing plasma shape diagnostic methods in the realm of real-time feedback control.

\vspace{0.3cm} 

\end{abstract}

%
%
%
%
\ioptwocol

\section{Introduction}

For magnetic confinement fusion devices like tokamak \cite{ferron1998real}, accurate measurement of plasma shape is one of the most preliminary steps toward subsequent diagnostics and control. Currently, most measurement techniques like magnetic equilibrium reconstruction code EFIT are largely contingent on magnetic field sensors like probes, the accuracy of which is largely affected by the applicability of magnetic sensing \cite{Hutchinson2002Principles}. Long plasma discharges can suffer from drift of the magnetic signals due to the integral nature of these measurements \cite{Donné2012Diagnostics}. Similarly, plasmas with a low plasma current and large distance to the magnetic pick-up coils (for example during the ramp-up and ramp-down phase of the discharge or ITER first plasma) may result in weak magnetic signals \cite{hommen2014real}. Meanwhile, under severe conditions like high-energy neutron radiation and high temperatures, the performance significantly deteriorates and could even completely malfunction due to heat shock and magnetic forces. Therefore, to satisfy the requirements in future extremely high-temperature environments like ITER, developing an alternative method for real-time, non-magnetic measurement is highly desired \cite{Santos2012Reflectometry}. 

Correspondingly, the optical plasma boundary
reconstruction for plasma position control sounds promising. For example, Hommen \textit{et al.} proposed to reconstruct the plasma boundary from dual camera-based, poloidal-view wavelength images, and attained high qualitative and quantitative agreement with EFIT in the MAST device \cite{hommen2010optical}. Subsequently, they implemented real-time plasma vertical displacement control in the TCV device \cite{hommen2014real}. However, the work \cite{hommen2010optical,hommen2014real} is restricted to the analysis and
reconstruction of plasma discharges of a single configuration,
the ``TCV standard shot'' \cite{hommen2014real}, since it relies on an appreciable distance between the plasma boundary and the first wall results in well-defined boundary features in the camera images, with little polluting light from reflections or plasma-wall interaction in the regions of interest (ROI). Ravensbergen \textit{et al.} extracted the optical plasma boundary and radiation front for detached divertor plasmas from multi-spectral imaging, and showed the possibilities to reliably detect the divertor leg and radiation front by lightweight image processing tools \cite{ravensbergen2020development}. But the magnetic field distribution on the divertor leg can affect the system's sensitivity \cite{ravensbergen2020development}. 
Luo \textit{et al.} utilized a Least Square-based method to delineate ROI in Charge-Coupled Device (CCD) images manually and demonstrated the successful reconstruction of plasma boundaries in EAST tokamak \cite{luo2018optical}. Nevertheless, the manual set of ROI makes these works \cite{hommen2014real,ravensbergen2020development,luo2018optical} less competent in automatically accomplishing the plasma shape inference task.

\begin{figure}[!tbp] 
\centering
\begin{subfigure}{.155\textwidth}
  \centering
  \includegraphics[width=.85\linewidth]{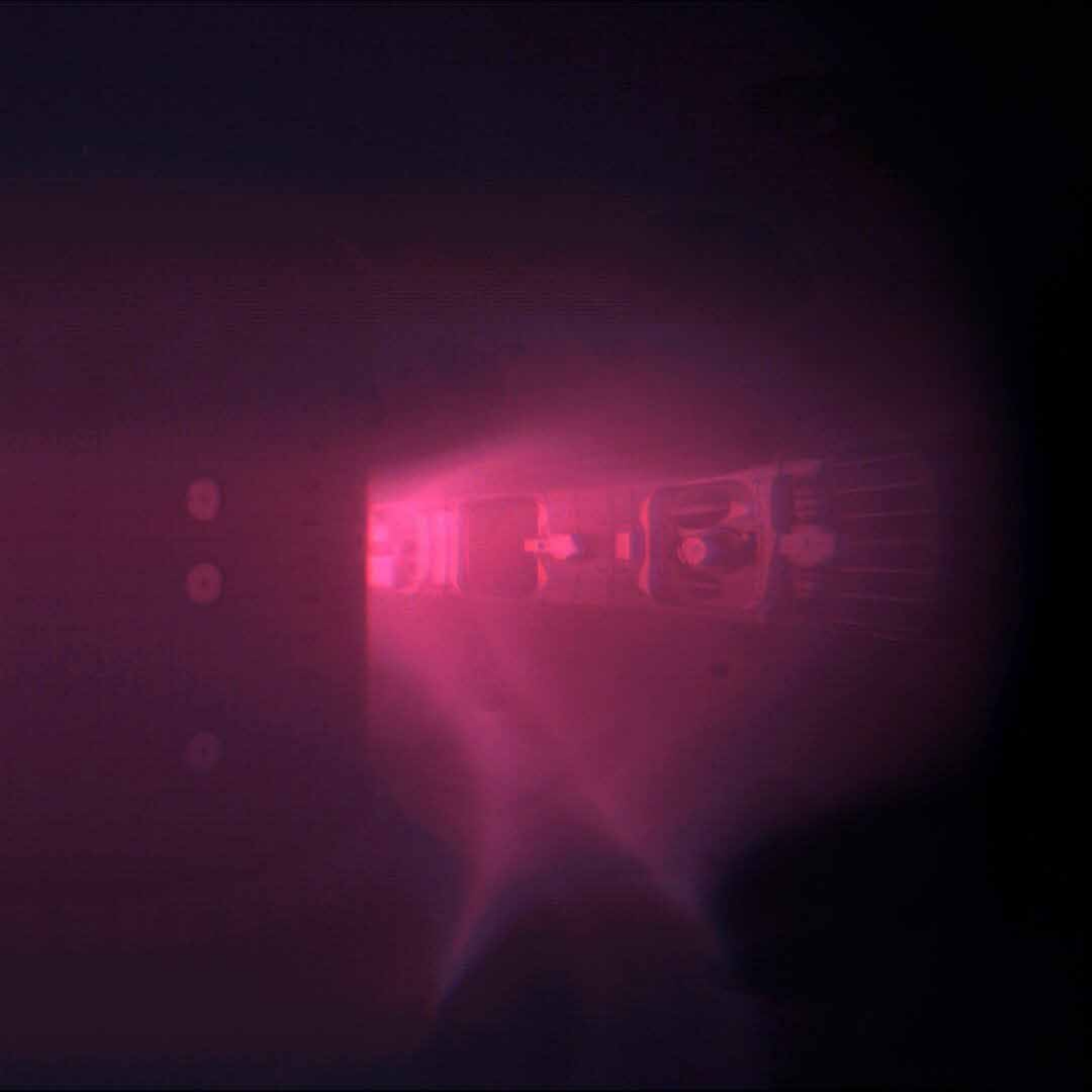} 
  \caption{} 
\end{subfigure}%
\begin{subfigure}{.155\textwidth}
  \centering
  \includegraphics[width=.85\linewidth]{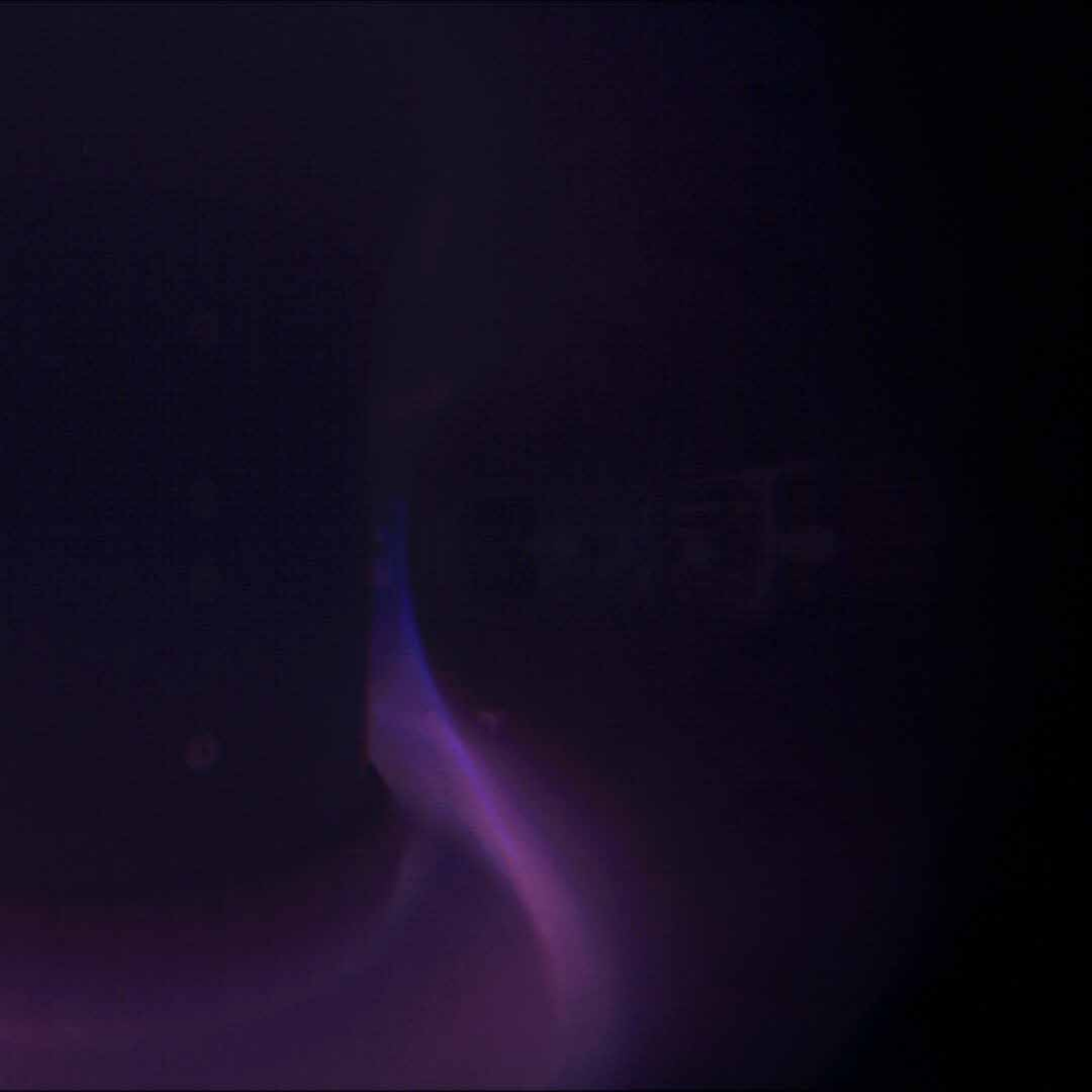}
  \caption{}
\end{subfigure}
\begin{subfigure}{.155\textwidth}
  \centering
  \includegraphics[width=.85\linewidth]{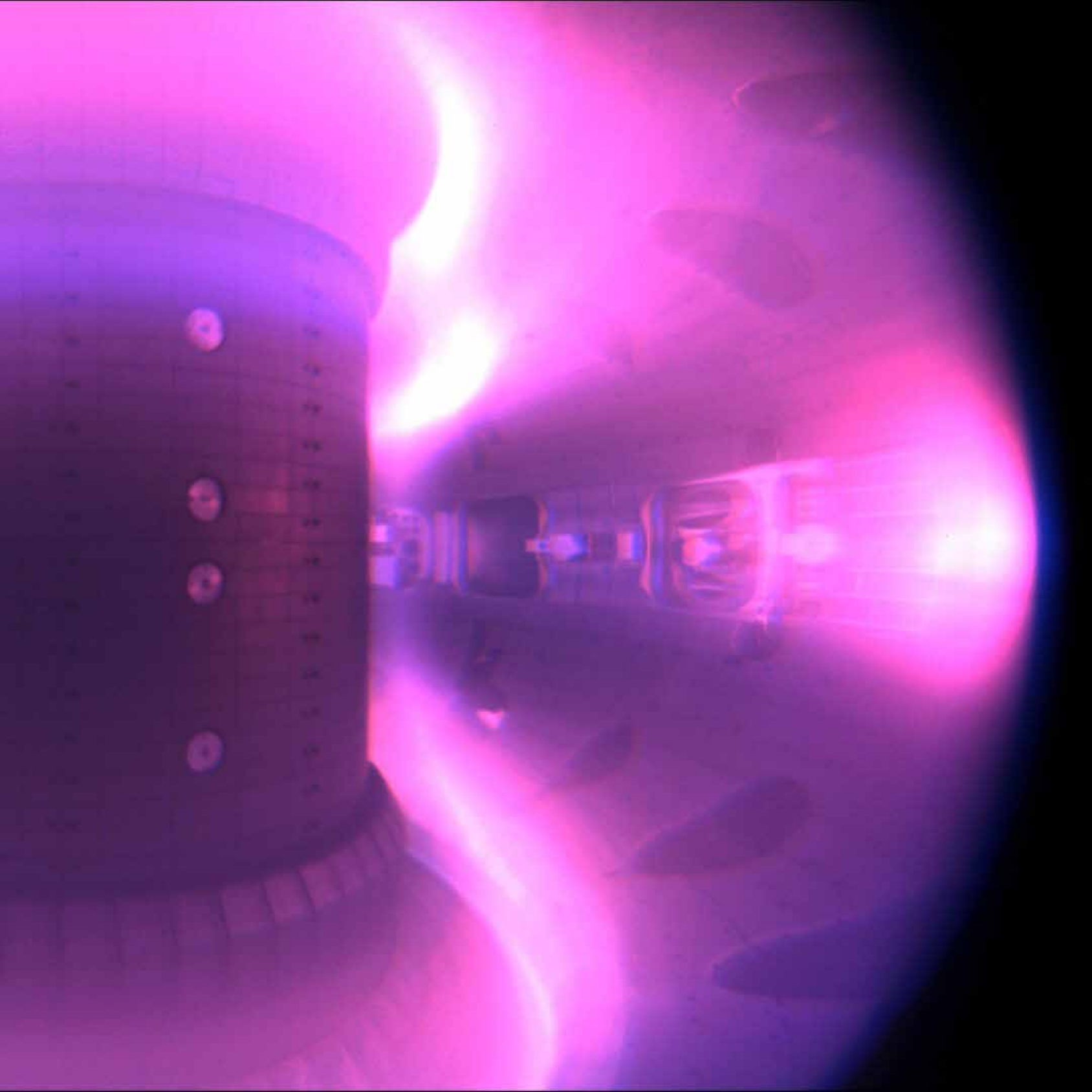}
  \caption{}
\end{subfigure}
\caption{Images of plasma shape captured with a camera on the HL-3 device. (a) displays the inner limiter shape in shot $\#06696$, with window interference present on the wall. (b) is the divertor shape related to shot $\#06696$, wherein the image exhibits low brightness levels, leading to a loss of edge recognition. (c) is a gas puffing interference image in shot $\#06255$, with bright air delivery interference spots visible on the wall.}
\label{fig:plasma}
\end{figure}

On the other hand, attributed to the powerful representation ability and efficient parallel computing, the application of machine learning (ML) has yielded significant progress in equilibrium reconstruction \cite{lao2022application,dong2021deep} and solver \cite{joung2019deep,cheng2021using}, plasma control \cite{degrave2022magnetic,seo2024avoiding} and boundary detection \cite{szHucs2022deep,yan2023optical}. In particular, by learning from labeled thermal patches on the initial plasma wall of the W7-X device, Szucs \textit{et al.} employed a vision-centric, YOLOv5-based method to infer possible thermal patchs while achieving near-real-time inference times \cite{szHucs2022deep}. Based on manually labeled plasma boundaries, Yan \textit{et al.} utilized the U-Net neural network to identify the boundary on EAST \cite{yan2023optical}. Besides suffering from manual labeling errors, the complexity of tokamak plasma poses another considerable challenge for accurate shape reconstruction. As shown in Figure \ref{fig:plasma}, factors like shifting plasma position, temperature, brightness, and current density distribution can impact the overall brightness within the device, thereby increasing the difficulty of predicting positional parameters. Additionally, the light spots caused by inner wall windows and gas injection interfere visually with plasma shape features, thus further adding to the reconstruction difficulty. Consequently, it is crucial to apply more powerful DNNs (e.g., Transformer \cite{vaswani_attention_2017}) for inference from CCD images of the complex environment.
Notably, though deep neural networks (DNN)-based surrogate models promise faster inference speed with satisfactory accuracy \cite{kates-harbeck_predicting_2019,kim_highest_2024}, it assumes the adoption of Multi-Layer Perceptron [MLP] and LSTM during the implementation. However, 
as the width and depth of DNNs escalate, the computational complexity of the Transformer neural network, the backbone of widely adopted image processing DNNs, increases triply \cite{vaswani_attention_2017}. Therefore, achieving real-time computation on cost-effective hardware platforms (e.g., Nvidia GeForce RTX 2080 Ti) becomes more challenging. In other words, for the real-time detection and control of plasma shape, it becomes imperative to strike the balance between accuracy and inference speed.

In this work, on top of Swin Transformer \cite{liu2021swin}, we develop a series of Poolformer \cite{yu2022metaformer, li2022efficientformer} Swin Transformer (PST) models, which demonstrate sufficient inference speed and relatively higher plasma reconstruction accuracy after training from large-scale data from the HL-3. In particular, we adopt a multi-task learning \cite{liebel2018auxiliary} framework by taking CCD images as the input while simultaneously utilizing the output of EFIT including radial position $R$ and vertical position $Z$ of plasma geometric center, minor radius $a$, elongation $\kappa$, upper triangularity $\delta_{u}$, and lower triangularity $\delta_{l}$ as labels. Such a design also avoids the cumbersomeness of manually labeling ROIs and contributes to learning consistency. To combat the unreliability of EFIT calculations during stages where the absolute value of plasma current is small but the rate of change is large, we incorporate a dynamic weight strategy as well \cite{kendall2018multi}. Furthermore, a knowledge distillation procedure is utilized for further compressing the model. The adapted Swin Transformer model manifests the comprehensive adaptability to the visual complexity of the HL-3 device plasma, including overly blurred boundaries, Neutral Beam Injection (NBI) and gas puffing interference, sudden bright spots, and wall hole interference. 
This specific model can seamlessly integrate with a plasma control system (PCS) and effectively support immediate magnetic field control.

\begin{table*}[ht]
\centering
\footnotesize
\caption{Input and Output of CCD-based Plasma Shape Reconstruction Model.}
\begin{tabular}{@{}ccccc}
\br
Input/Output & Variable & Unit & Dimension & Description \\
\mr
Input & Image & N/A & $t\times 3\times 120\times 120$ & Input image for the CCD perception model. \\
Input & Exposure Time & ms & $1$ & Interval time of image acquisition by the camera. \\
Output & $R$ & cm & $1$ & Radial position of plasma geometric center \\
Output & $Z$ & cm & $1$ & Vertical position of plasma geometric center \\
Output & $a$ & cm & $1$ & Minor radius \\
Output & $\kappa$ & dimensionless & $1$ & Elongation \\
Output & $\delta_{u}$ & dimensionless & $1$ & Upper triangularity \\
Output & $\delta_{l}$ & dimensionless & $1$ & Lower triangularity \\
\br
\end{tabular}
\label{table_inputoutput}
\end{table*}

\section{Methods}
\subsection{Data Collection and Pre-Processing}
\label{sec:data_processing}
HL-3 is a medium size tokamak with an aspect ratio of $2.8$: plasma current $I_{p}$ = $2.5$–$3$ MA, toroidal field $B$ = $2.2$–$3$ T, major radius $R$ = $1.78$ m, minor radius $a$ = $0.65$ m, and elongation $\kappa\leq1.8$; triangularity $\delta\leq0.5$\cite{duan2022progress}. HL-3 was designed to have a flexible configuration in order to explore multiple divertor configurations. Three HCD systems are able to provide a total power of $27$ MW, including $15$ MW of NBI, $8$ MW of electron cyclotron resonance frequency (ECRF), and $4$ MW of LHCD. In this work, given the clear evidence of the positive relationship between the CCD image and EFIT of a plasma shape \cite{luo2018optical,yan2023optical}, we aim to directly learn the characteristics of a plasma CCD image and reconstruct the plasma shape according to the output of EFIT. In other words, the input is the CCD image taken from HL-3, while the output is the corresponding parameters (i.e., $R$, $Z$, $a$, $\kappa$, $\delta_{u}$, and $\delta_{l}$) for the plasma shape output by EFIT. Table \ref{table_inputoutput} summarizes the details of the input and output. 

Specifically, the HL-3 device is equipped with an advanced visible light diagnostic system, providing three disparate views - tangential, extreme wide-angle, and downward-looking - of the dynamics of each discharge. The tangential view theoretically provides clearer information on the upper and lower divertor geometry, enhancing the capture of lateral plasma behavior. Yet, the proximity of the tangential lens to the heating area on HL-3 often leads to lens coating, resulting in over-blurred lens imagery, limited effective data, and complications in plasma observation. Consequently, we primarily employ CCD images from the extreme wide-angle view due to their robustness. Albeit its sub-optimality, our results demonstrate that it still yields fairly accurate reconstruction results. We believe that once the engineering issues are resolved, the tangential view with richer information on vertical displacements promises further performance improvement.

\begin{figure}[!tbp] 
\centering
\includegraphics[width=.85\linewidth]{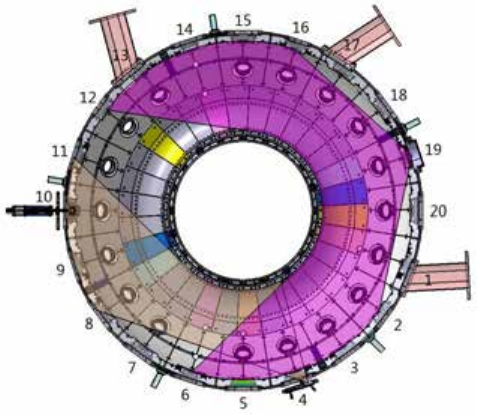}
\caption{Toroidal cross section of the HL-3 tokamak \cite{liu2024multidirectional}, including the tangential view diagnostic (orange color, from $\#4$ equatorial port), the extreme wide-angle view diagnostic (magenta color, from $\#19$ equatorial port), and the downward-looking view diagnostic installed in the $\#5$ sector, which is indicated by a green arrow. The location of the gas puffing entry at $\#10$ equatorial port is indicated by a red arrow. In this study, we use the extreme wide-angle view diagnostic.}
\label{fig:viewing}
\end{figure}

The CCD camera for the extreme wide-angle view outputs images in the Bayer GB8/Bayer GB10 format, with a high frame rate of up to $2,000$ fps and a resolution of $1,920 \times 1,080$ pixels, each measuring $10\ \mu m \times 10\ \mu m$. This high frame rate ensures the capture of subtle dynamic changes in plasma, vividly reflecting the morphology and dynamic changes of the plasma, and providing abundant visual information for diagnosis. Additionally, the Bayer algorithm can convert these images into an RGB format without loss of quality. Notably, the exposure time is taken into account as well, given the inherent brightness differences caused by adjustments of exposure time (i.e., $1 \sim 2$ ms) from the camera API.

During the experiment, we collect a dataset of CCD images with exposure time information from Shot $\#05554$ to $\#06226$. In particular, we select $271$ effective shots, characterized by the plateau phase lasting a minimum of $500$ ms and a consistent plasma current of $100$ kA during the phase. This selected dataset corresponds to a combination of $324,911$ images. For each image, it undergoes initial cropping followed by linear interpolation to achieve a size of $120 \times 120$ pixels. Afterward, the compressed images are subject to a group normalization operation, aiming to mitigate discrepancies in brightness and contrast among them. Meanwhile, no denoising operation is conducted, as this helps to effectively retain the nuanced contrasts within the images. 
Among the selected shots, $236$ are designated for training, with the remaining $35$ allocated for testing. Given the typically high similarity between adjacent shots, we meticulously adjust the dataset division to guarantee the existence of shots with noticeable discrepancies in reference templates and plasma shape under both the training and testing datasets. This precautionary measure contributes to preventing the model overfitting.

As mentioned earlier, EFIT might encounter convergence issues for certain cases, which is difficult to detect and could cause significant interference with the training of the CCD model. Therefore, we use EFITNN \cite{joung2019deep,lao2022application}, a DNN surrogate model of EFIT, which is capable of obtaining the boundary shape of the plasma. We use six primary parameter EFITNN outputs ($R$, $Z$, $a$, $\kappa$, $\delta_{u}$, and $\delta_{l}$) as labels in our model. For each parameter, the actual maximum and minimum are extracted from effective shots' statistics, and a min-max normalization operation, which brings any parameter $x \in \{R, Z, a, \kappa, \delta_{u}, \delta_{l}\}$ to an interval of $(0,1)$ as $x \leftarrow \frac{x- \min}{\max - \min}$, is applied accordingly. We find that such an operation can prevent potential biases in the learning process.

\subsection{ML Model}

We initially consider following the DNN structure of Swin Transformer \cite{liu2021swin} to learn the relationship between CCD images and plasma shape parameters ($R$, $Z$, $a$, $\kappa$, $\delta_{u}$, and $\delta_{l}$). Specifically, to connect an image to the output, Swin Transformer \cite{liu2021swin} adopts a parallelizable multi-head attention mechanism in Transformer \cite{vaswani_attention_2017, niu2021review}, which completely dispenses recurrence and convolutions. Swin Transformer then employs a Convolutional Neural Network (CNN)-alike methodology to extract the hierarchical feature maps of a CCD image. Generally, Swin Transformer consists of $4$ stages, each beginning with a patch merging and layer normalization operation to gradually downsample the image. Intuitively, as for an RGB image of $3\times 120\times 120$ (indicated as $3\times H \times W$), Swin Transformer changes it to a size of $C\times\frac{H}{4}\times\frac{W}{4}$, $2C\times\frac{H}{8}\times\frac{W}{8}$, $4C\times\frac{H}{16}\times\frac{W}{16}$, $8C\times\frac{H}{32}\times\frac{W}{32}$ after each stage. Notably, to maintain an acceptable computational cost, the Swin Transformer does not directly calculate the global attention from the entire image. Instead, it partitions the image into fixed-size windows (in our study, we adopt a window size of $5\times5$), and employs a Windows-Multi-head Self-Attention (W-MSA) mechanism \cite{liu2021swin} for each window. However, the lack of inter-window information exchange confines the model to capture the long-range relationship. Therefore, a Shifted Window-MSA (SW-MSA) block, which shifts each window half of the window size downward and rightward respectively, enhances the performance. In practice, Swin Transformer alternately uses W-MSA and SW-MSA block. Detailed implementation procedures are provided in the Appendix. 

Swin Transformer offers multiple versions, including Swin-tiny, Swin-small, Swin-base and Swin-Large, each increasing in model size and complexity. We utilize the simplest and lightest one, Swin-tiny, with a depth (i.e., no. of blocks) of $2$, $2$, $6$, and $2$ at each stage. As discussed lately, Swin Transformer can accurately interpret local details and global brightness distribution of the plasma, producing precise parameter predictions even under a wide and variable range of plasma conditions. Nevertheless, even for the smallest Swin-tiny model, it takes at least $10$ ms to perform the inference on Nvidia GeForce RTX 2080 Ti, which hinders its applications in high-frequency real-time magnetic control.

\begin{figure}[!tbp] 
\centering
\includegraphics[width=\linewidth]{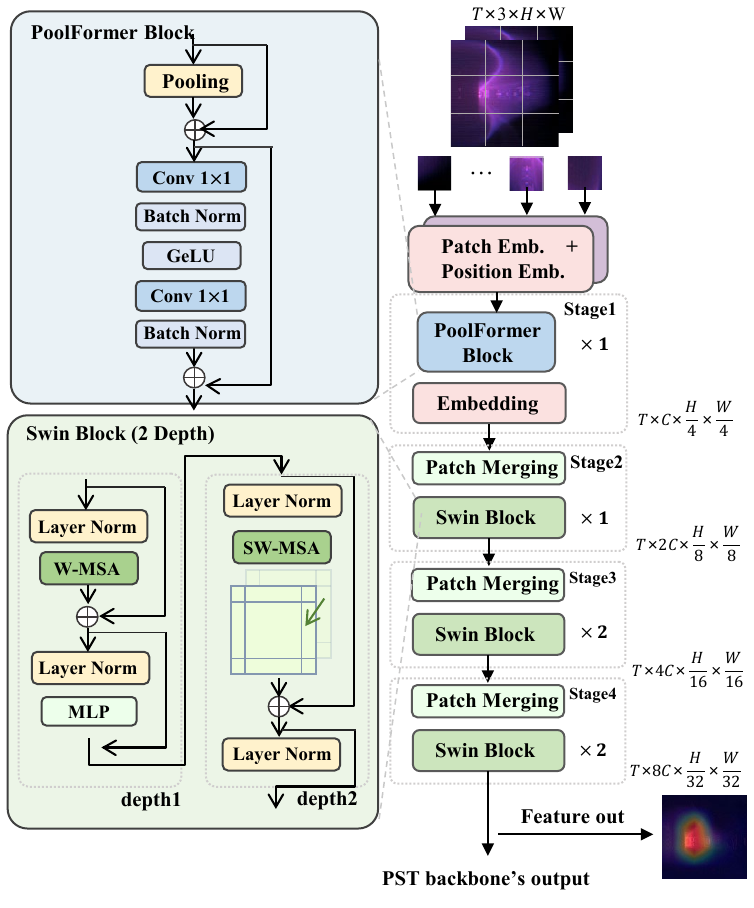}
\caption{Illustration of the shared base DNN of PST.}
\label{fig:Fundamental network}
\end{figure}

To address this issue, we take inspiration from \cite{yu2022metaformer, li2022efficientformer} and develop a PST-tiny DNN. Particularly, PST-tiny only keeps a portion of blocks (i.e., the first $1$, $1$, $2$, and $2$ blocks) for four stages, and reduces the dimension of the hidden layer from $64$ in Swin-tiny to $32$. More specifically, the Swin block in the first stage is replaced with a more computation-efficient Poolformer block \cite{yu2022metaformer, li2022efficientformer}. In particular, as its name implies, Poolformer divides the input tokens (i.e., partitioned images) into multiple groups and then performs a pooling operation within each group. Theoretically, for the number of input tokens $n$, MSA has a computational complexity of $O(n^2)$, while Poolformer only needs $O(n)$ computations \cite{yu2022metaformer, li2022efficientformer}.

In addition, we leverage convolution and batch normalization operations rather than layer normalization in Swin Transformer, given their typically faster speed and easier parallelizable implementation on modern hardware.  Notably, our experimental results show that the DNN structure re-design can contribute to a reduction in inference time of over $50\%$ compared to the Swin Transformer. 
As a comparison, we name the Swin-tiny with an added Poolformer as PST-base.

\subsection{Model Training}
In this part, we first outline the design of the loss functions. Specifically, we leverage a multi-task learning loss with a dynamic weight strategy to efficiently learn the six plasma shape parameters, and adopt a knowledge distillation loss to further compress the model. Following this, we detail the practical training procedures.
\subsubsection{Loss Function Design} 
\paragraph{(a) Individual task loss}
In response to the observed difficulty of the EFIT solution to achieve convergence during the ramp-up and ramp-down phases of plasma discharge, we opt to implement a particular loss function. In this context, given the potential outliers for $R$, $Z$, and $a$ calculated during the ramp-up and ramp-down phases, we choose the Huber loss function, which combines Mean Squared Error (MSE) and Mean Absolute Error (MAE), to effectively mitigate the negative impact. Mathematically, for $x \in \{R, Z, a\}$,
\begin{align}
L_{\text{Huber}}(x, x') = 
  \begin{cases} 
    \frac{1}{2}(x - x')^2,  & |x - x'| \leq \delta; \\
    \delta |x - x'| - \frac{1}{2}\delta^2, & \text{otherwise},
  \end{cases}\label{eq:huber_loss}
\end{align}
where $x'$ is the output by the PST model corresponding to a CCD image while $x$ denotes the corresponding EFITNN-output label. Besides, the hyperparameter $\delta$ is set as $1$. Meanwhile, for the parameters $\kappa$, $\delta_u$ and $\delta_d$, which have a certain tolerance for prediction errors, we use the MAE loss function $L_{\text{MAE}}(x, x') = |x - x'|$ exclusively to avoid over-penalization. The corresponding formula for a batch can be summarized as
\begin{align}
R(x) = 
  \begin{cases} 
    \frac{1}{N} \sum\nolimits_{i=1}^{N} L_{\text{Huber}}(x_{i}, x'_{i}),  & x = R, Z, a; \\
    \frac{1}{N} \sum\nolimits_{i=1}^{N} L_{\text{MAE}}(x_{i}, x'_{i}), & x = \kappa, \delta_u,\delta_d\,
  \end{cases}\label{eq:R(x)}
\end{align}
where $N$ represents the batch size. 

\paragraph{(b) Multi-Task Learning Loss}

\begin{figure}[!tbp] 
\centering
\includegraphics[width=0.70\linewidth]{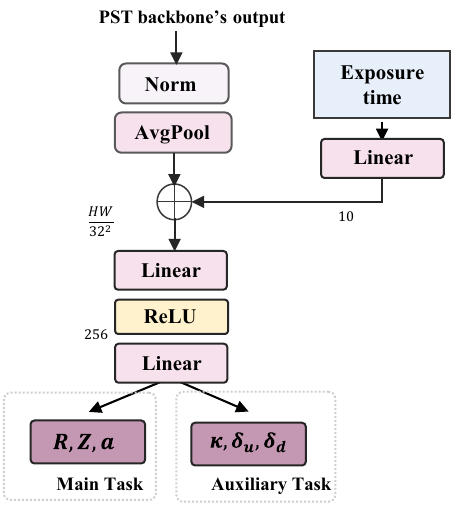}
\caption{Final DNN part of PST for multi-task learning framework.}
\label{fig:multi_task}
\end{figure}

Since $R$, $Z$ and $a$ in a CCD image exhibit more intuitively distinctions than $\kappa$, $\delta_u$, and $\delta_d$, and the PCS in HL-3 is contingent on $R$, $Z$, and plasma current $I_p$ for control, we perform a Multi-Task Learning (MTL) by treating the learning of $R$, $Z$, and $a$ as main tasks while regarding the perception of $\kappa$, $\delta_u$, and $\delta_d$ as auxiliary tasks. Such an MTL design contributes to learning the similarities across tasks, while
the incorporation of auxiliary tasks allows for more effective optimization of the DNN parameters, thereby alleviating overfitting.
Although the previous study \cite{kendall2018multi} has shown that jointly trained DNNs outperform those trained separately for each task, determining appropriate MTL weights is costly, especially when multiple plasma shape parameters are required. Therefore, 
to effectively balance the individual contribution of the six shape parameters, we employ a multi-task loss strategy \cite{kendall2018multi}. 

Accordingly, the loss function for a batch in MTL can be written as

\begin{align}
L_{\text{MTL}}(x) &=  \sum\nolimits_{x \in \{R, Z, a\}} \left( \frac{1}{2c_x^2} R(x) + \ln (1 + c_x^2) \right) \nonumber \\
& \ + \sum\nolimits_{x \in \{\kappa, \delta_u,\delta_d\}} \alpha\left( \frac{1}{2c_x^2} R(x) + \ln (1 + c_x^2) \right),\label{eq:MTL}
\end{align}
where $\ln(1+c_x^2)$ is applied as a regularization term, and the learnable network parameter $c_x,\forall x \in \{R, Z, a, \kappa, \delta_{u}, \delta_{l}\}$ contributes to automatically accounting for the different variances and biases among single-task losses. This approach effectively balances the loss weights of the six parameters, thereby achieving superior performance. 
During practical training, the three parameters in the auxiliary tasks, which receive less attention, are initialized as smaller values ($0.5$, $0.5$, $0.5$), whereas the initial values for the parameters in the main tasks are set as $1$.
The hyperparameter $\alpha$ is used for the auxiliary tasks to prevent them from becoming overly important, and it is set at 0.2 in our study.

\paragraph{(c) Knowledge Distillation Loss for Model Compression}

\begin{figure}[!tbp] 
\centering
\includegraphics[width=0.85\linewidth]{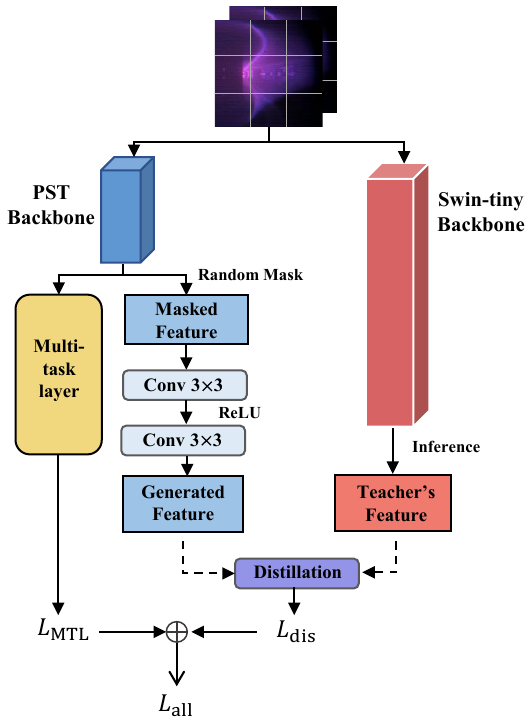}
\caption{Illustration of the Masked Generative
Distillation (MGD) structure. The student's feature is randomly masked and then the projector layer is utilized to induce the student to generate the teacher's feature using the masked one.}
\label{fig:distillation}
\end{figure}
Despite the enhanced time efficiency of PST-tiny, it exhibits a significant accuracy gap compared to PST-base model. In order to compensate for the performance loss, we employ the method of knowledge distillation, by using PST-tiny as a more streamlined student DNN model and PST-base as the teacher model. This modification allows PST-tiny to deliver more reliable results while enhancing its efficiency.

Concretely, the plasma shape reconstruction task falls into the scope of regression learning, where conventional label-oriented distillation techniques struggle to generate consistent continuous output.  Consequently, we turn to feature-based knowledge distillation techniques and adopt Masked Generative Distillation (MGD) \cite{mgd}, which uses features (i.e., the hidden layer output) demonstrated by a well-trained teacher DNN to guide the learning of the student DNN. Furthermore, MGD applies random masking to pixels of the student's feature and forces it to generate the teacher's full feature through a simple adaptation layer. In MGD, random pixels are used in each iteration, ensuring all pixels are eventually utilized throughout the training process. Meanwhile, the distillation occurs at the final layer of the Swin Transformer blocks in both networks with MSE computed for pixel values at corresponding positions on the feature maps. Mathematically, for a feature map with size $C\times H\times W$,

\begin{align}
L_{\text{dis}} = \sum_{k,i,j=1}^{C,H,W} \left(F^\mathbf{T}_{k, i, j} 
 - G(f_{\text{align}}(F^\mathbf{S}_{k, i, j}) \times M_{k, i, j})\right)^2, 
\end{align}
where the superscripts $\mathbf{S}$ and $\mathbf{T}$ correspond to the student and teacher DNNs, respectively. $f_{\text{align}}$ represents the adaptive layer that aligns student features with teacher features, while the student's feature map is randomly masked according to the masking matrix $M$. 
$G= W_{l2}(\text{ReLU}(W_{l1}(F)))$, which incorporates two convolution layers ($W_{l1}$, $W_{l2}$) and one activation layer ($\text{ReLU}$), represents the projector layer and leads to consistent dimensions of feature maps in student and teacher models.

Since the teacher DNN only serves as a guide for student DNNs to restore features rather than requiring direct imitation, it effectively enhances the capability of the student DNN to generate and understand complex structural data.
\subsubsection{Training Procedures}
For each image with a particular exposure time, six corresponding plasma shape parameters are used as labels. Considering the unreliability of EFIT calculations during periods where the absolute value of plasma current $I_p$ is small but the rate of change is large, owing to factors like eddy currents, we choose to dynamically adjust the confidence level of the labels. For each set of training data, we regard the first $150$ ms as the ramp-up period and the last $150$ ms as the ramp-down period.  Accordingly, given our limited confidence in EFIT-output labels in these two periods, we assign smaller weights when calculating the loss function. On the contrary, due to the high confidence in the label accuracy during the plateau stage, we give it a higher weight. Such a dynamic weight strategy guides the model to focus more on improving performance during the plateau stage, which indirectly enhances its efficacy in the other phases. Accordingly, the MTL loss function is updated as
\begin{align}
& L_{\text{MTL}} = \gamma \cdot L_{\text{MTL}}(x)
\end{align}
where the parameter $\gamma$, which is associated with the specific moment $t$ within a shot, can be written as 
\begin{align}
\gamma  = 
\begin{cases} 
0.8, & \text{if } t < 150 \text{ or } t > t_{\max} - 150, \\
1, & \text{otherwise.} 
\end{cases}
\end{align} 
Here $t_{\max}$ refers to the moment at which the discharge ends.

On the other hand, by jointly involving the MTL loss and knowledge distillation loss, the combined loss function can be formulated as
\begin{align}
L_{\text{comb}} = (1-\alpha)L_{\text{MTL}} + \alpha L_{\text{dis}}, 
\end{align}
where $ \alpha $ is a hyperparameter used for balance loss. Once the magnitudes of the original loss and feature loss are unified, we set $\alpha=0.1$ for our experiments. This approach successfully ensures the performance of the PST-tiny model while significantly improving its efficiency.

\subsection{Model Deployment}
Given a well-trained PST-tiny model, we utilize ONNX to convert it from the PyTorch model saving format (.pth) to the more universal ONNX format. This open and widespread model ONNX format facilitates convenient model sharing across multiple deep learning frameworks. Additionally, we apply the ONNX Simplifier tool, which capably infers the entire computation graph and replaces redundant operators with constant outputs, to simplify the model and enhance inference speed.

Ultimately, our model has been successfully deployed on a Windows operating system, with model inference running on Nvidia GeForce RTX 2080 Ti. Notably, we apply the powerful deep learning inference optimizer and runtime environment, Nvidia TensorRT \cite{xia2022trt}, for subsequent optimization of models in the ONNX format. Typically, the model optimization, which involves layer fusion, operator fusion, model pruning, and precision calibration, further increases the model run efficiency and boosts the model performance. Our offline results suggest that the optimization contributes to further reducing half of the inference time. Table \ref{table:model_inference_comparison} summarizes the inference and deployment time. Meanwhile, as discussed lately, the model compression trivially affects the inference performance with less than $0.01$ cm compared to the PST-base model.

\begin{table}[t]
\centering
\footnotesize
\caption{Comparison of Inference Speed and Deployment Time for Different Models Tested on Nvidia GeForce RTX 2080 Ti.}
\label{table:model_inference_comparison}
\begin{tabular}{@{}>{\centering\arraybackslash}m{1.6cm}>{\centering\arraybackslash}m{1cm}>{\centering\arraybackslash}m{1.5cm}>{\centering\arraybackslash}m{1.5cm}@{}}
\br
Model & Depth & Inference Time (ms) & Deployment Time (ms) \\
\mr
Swin-tiny & $(2,2,6,2)$ & $3.4$ & $4.3$ \\
PST-tiny & $(1,1,2,2)$ & $1.0$ & $1.8$ \\
\br
\end{tabular}
\end{table}


\begin{figure*}[!ht]
    \centering
    \begin{subfigure}[b]{0.4\textwidth}  
        \includegraphics[width=1\columnwidth]{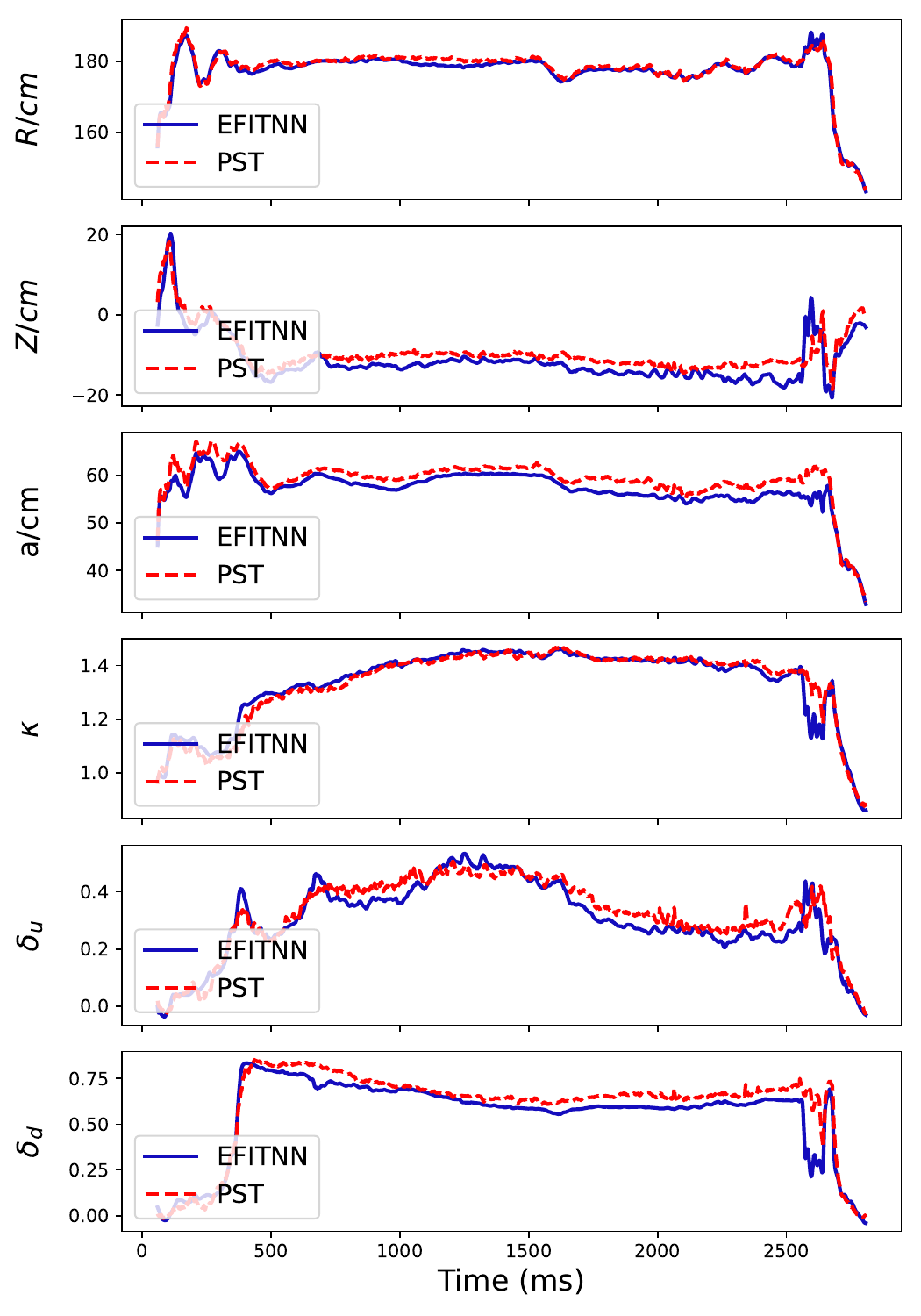}
        \caption{}
    \end{subfigure}
    \hspace{5mm}
    \begin{subfigure}[b]{0.4\textwidth}
        \includegraphics[width=1\columnwidth]{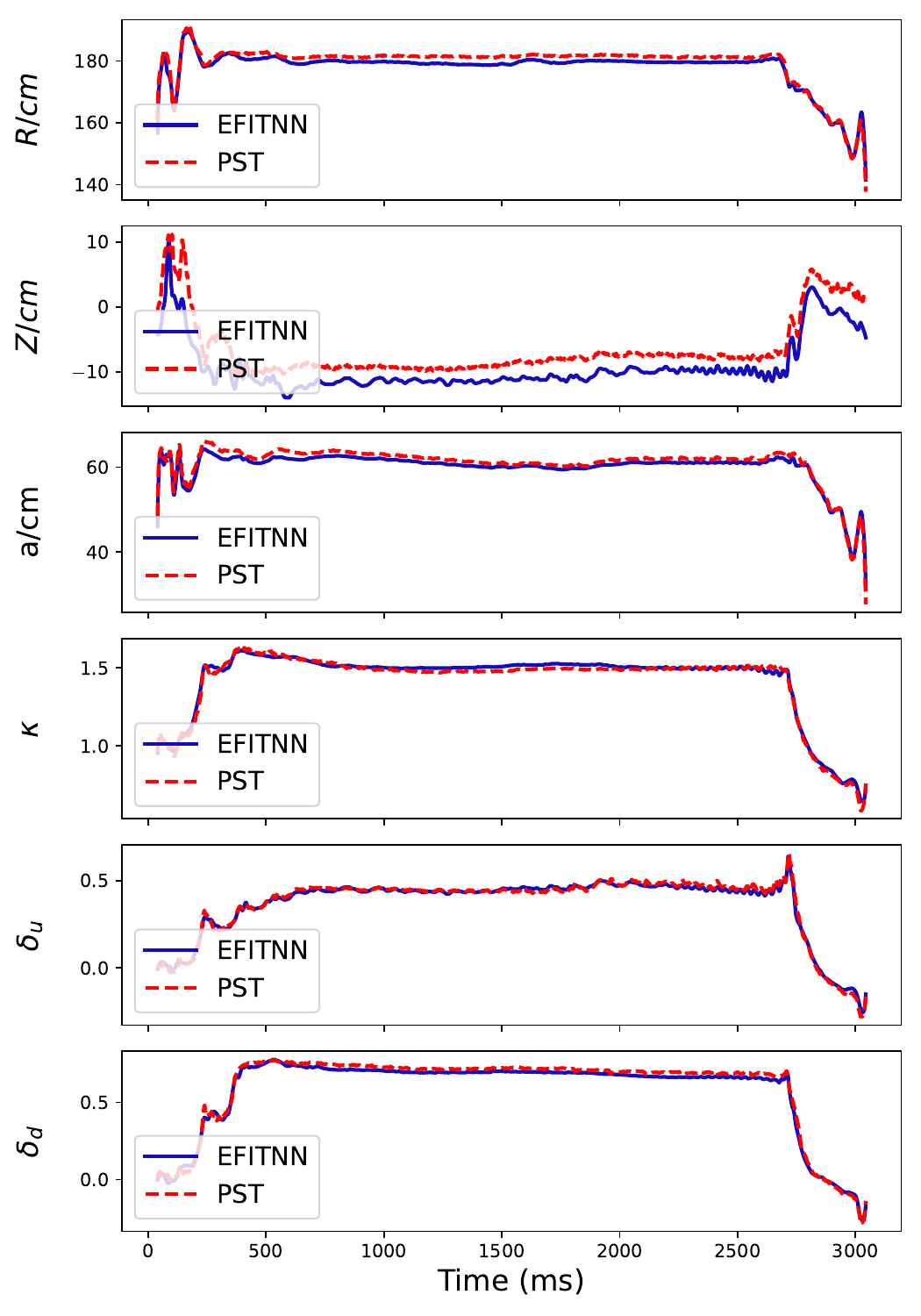}
        \caption{}
    \end{subfigure}
\caption{The comparison between the PST-based plasma shape detection results and EFITNN output on (a) \#06227 shot and (b) \#06236 shot.}
\label{fig:offline_results}
\end{figure*}

\begin{figure*}[!ht]
\centering
\includegraphics[width=\textwidth]{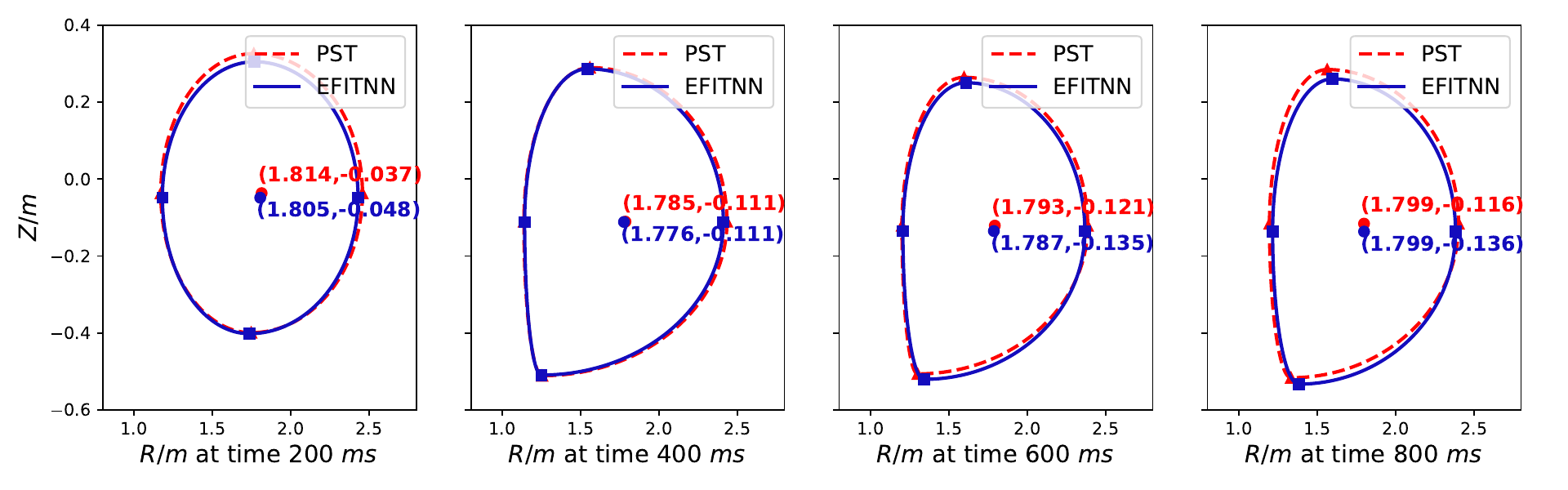}
\caption{Comparison of plasma boundary coordinates for shot $\#06227$ reconstructed using six parameters from the PST model and those using EFITNN.}
\label{fig:gridplot}
\end{figure*}

\begin{figure}[!tb]
    \centering
    
    \begin{subfigure}{.25\textwidth}
        \centering
        \vspace*{\fill}
        \includegraphics[width=1.1\linewidth]{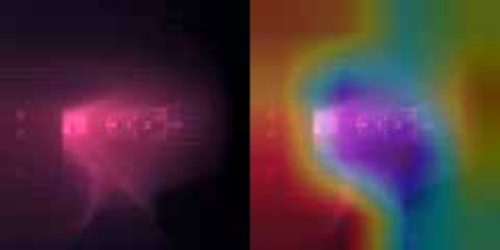}
        \vspace{\fill}
        \caption{}\label{fig:features_a}

    \end{subfigure}%
        \begin{subfigure}{.25\textwidth}
        \centering
        \vspace*{\fill}
        \includegraphics[width=.7\linewidth]{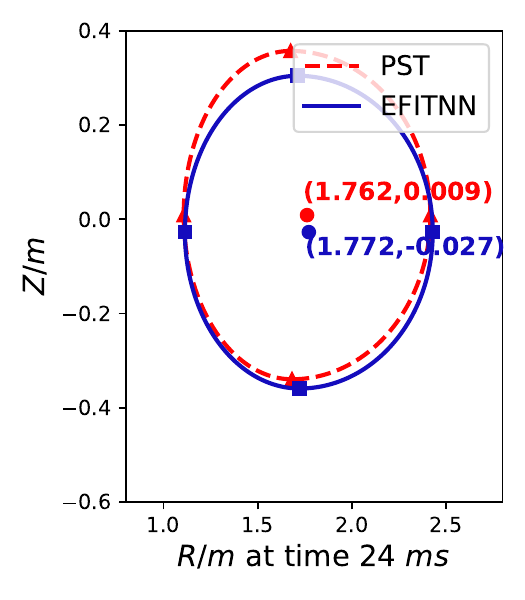}
        \vspace{\fill}
        
        \caption{}
    \end{subfigure}%
    
    \begin{subfigure}{.25\textwidth}
        \centering
        \vspace*{\fill}

        \includegraphics[width=1.1\linewidth]{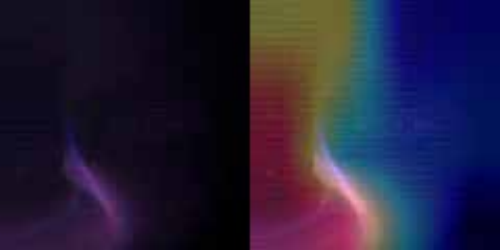}
        \vspace{\fill}
        \caption{}\label{fig:features_c}
    \end{subfigure}%
        \begin{subfigure}{.25\textwidth}
        \centering
        \vspace*{\fill}

        \includegraphics[width=.7\linewidth]{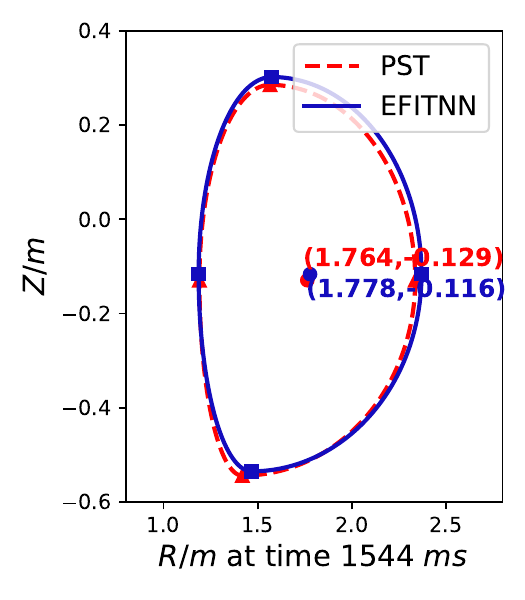}
        \vspace{\fill}
        \caption{}
    \end{subfigure}%
    
    \begin{subfigure}{.25\textwidth}
        \centering
        \vspace*{\fill}
        \includegraphics[width=1.1\linewidth]{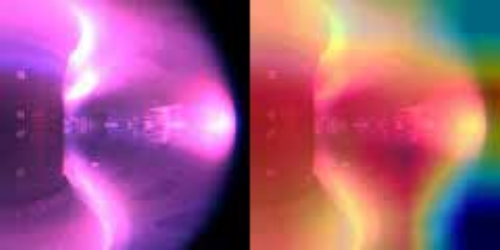}
        \vspace{\fill}
        \caption{}\label{fig:features_e}
        
    \end{subfigure}%
        \begin{subfigure}{.25\textwidth}
        \centering
        \vspace*{\fill}
        \includegraphics[width=.7\linewidth]{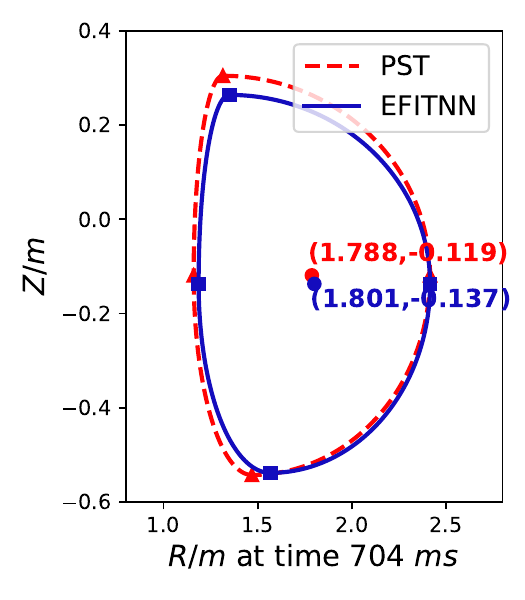}
        \vspace{\fill}
        \caption{}
    \end{subfigure}%
\caption{Visualization of feature maps and boundary detection results. Subfigures (a), (c), and (e) respectively represent the feature maps of the plasma under the states of inner limiter configuration, biased limiter configuration, and gas puffing process in Figure \ref{fig:plasma}, while subfigures (b), (d), and (f) provide the inferred boundary. Notably, the transition from blue to red in feature maps indicates increased attention or DNN weights to the related pixels during the computation.}
\label{fig:features}
\end{figure}

\begin{figure}[!ht] 
\centering
\includegraphics[width=0.9\linewidth]{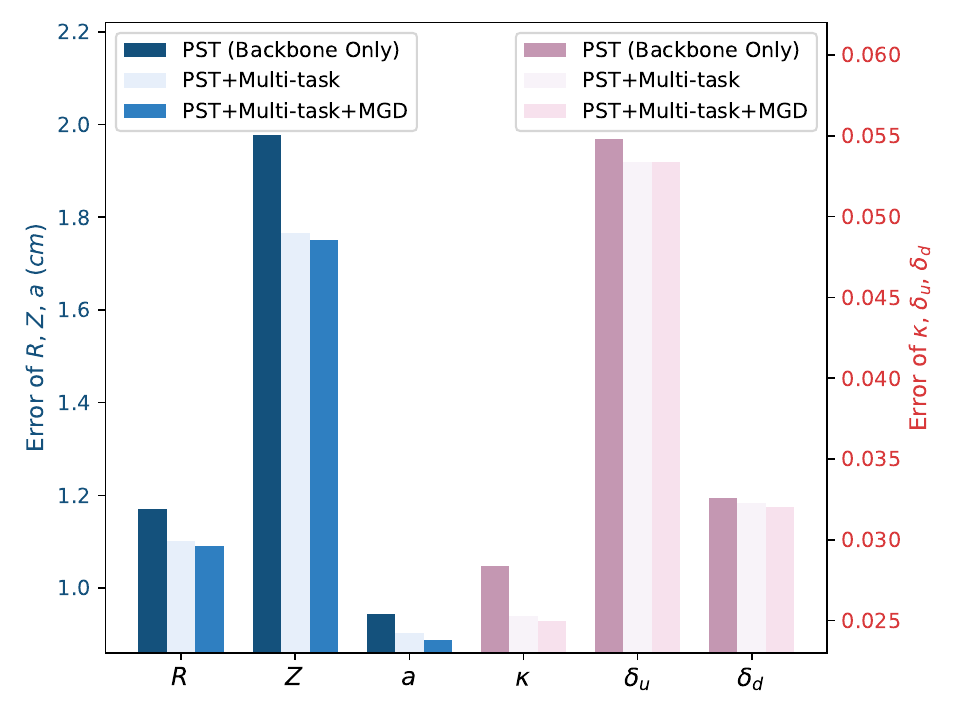}
\caption{Bar chart of the PST model's gain after adding different modules.}
\label{pdf:Module Gain}
\end{figure}
\section{Results}
\subsection{Offline Results}
In this part, we investigate the offline performance of DNN-based plasma shape detection results and evaluate their accuracy in comparison with EFITNN. 
As mentioned in Section \ref{sec:data_processing}, the testing dataset encompasses $35$ complete discharge shots. To ensure the evaluation completeness, the dataset is inclusive of the ramp-up, plateau, and ramp-down phases, and comprises an even distribution of plateau phases reaching $300$ kA and $500$ kA.

Figure \ref{fig:offline_results} illustrates the inferred six shape parameters by the PST model throughout the entire discharge process for shots $\#06227$ and $\#06236$, each reaching $300$ kA and $500$ kA during the flat-top period, respectively. Generally, the model can qualitatively capture subtle shape changes in key parameters, and the prediction accuracy of $R$ and $a$ is particularly impressive, with an MAE margin within $1.5$ cm. However, the accuracy for $Z$ is slightly lower with approximately $1.8$ cm on average. Correspondingly, $\kappa$, $\delta_{u}$, and $\delta_{l}$, all highly correlated with vertical displacement $Z$, exhibit discrepancies in some cases.

Notably, such inferiority aligns with the adoption of the extreme wide-angle view, as opposed to the tangential view, which might miss some important changes in the upper and lower divertor. 
Furthermore, we try to explore the potential of spatio-temporal fusion techniques to mitigate image noise errors, by stacking $T$ consecutive images. Although this approach promises enhanced accuracy, its lengthy inference and deployment time hinders practical application.

To maintain a clear understanding of the plasma shape, we compute the extreme points of the four sections from detected shape parameters and then approximate the plasma boundary as the linkage of four quarter-ellipses. Figure \ref{fig:gridplot} presents the corresponding result for shot $\#06227$, and demonstrates the proficient mimicking of the plasma's progression towards the divertor shape. On the other hand, for cases in Figure \ref{fig:plasma}, the left part of Figure \ref{fig:features} illustrates the feature maps, which are attained by extracting the intermediate outputs from the shared base DNN of the model in Figure \ref{fig:Fundamental network}. Overall, the color in the feature map reflects the importance of individual pixels for subsequent computations, and the transition from blue to red indicates increased attention (i.e., larger DNN weights) to related pixels. Despite the consistently high attention on RoIs, 
subtle differences can still be observed in local feature maps. For internal limiter plasmas with lower plasma current $I_p$ in Figure \ref{fig:features_a}, the feature map distinctly outlines plasma shape boundary and likely contributes to a more defined and easily identifiable boundary. For the divertor-shaped plasmas shown in Figure \ref{fig:features_c}, the feature map perdominantly focuses on the pixel points at the divertor tagret plate, thereby enhancing the spatial positioning of the divertor configuration. Meanwhile, for gas puffing in Figure \ref{fig:features_e}, where the gas injection point overlaps with the outer boundary of the plasma, the presence of extra bright spots severely interferes with the boundary recognition. Consequently, the feature map puts more emphasis on the changes in brightness inside the plasma. Correspondingly, the recovered plasma boundary shown in the right part of Figure \ref{fig:features} demonstrates high consistency with EFITNN in these challenging cases. Meanwhile, we investigate the gains of modular design in PST and provide the results in Figure \ref{pdf:Module Gain}. It can be observed that additional enhancements in model performance can be anticipated after the integration of MTL and MGD techniques, especially in terms of the vertical displacement $Z$.

\begin{table*}[!ht]
\centering
\footnotesize
\caption{Comparison of MAE for Plasma Shape Parameters Detected by Different Image Perception Models.}
\label{table:offline_comparison}
\begin{tabular}{@{}ccccccccc}
\br
Model & Depth & R (cm) & Z (cm) & a (cm) & $\kappa$ & $\delta_{u}$  & $\delta_{l}$ \\
\mr
PST-tiny & $(1,1,2,2)$ & $1.091$ & $\mathbf{1.752}$ & $0.887$ & $0.025$ & $0.053$ & $0.032$ \\
PST-base & $(2,2,6,2)$ & $\mathbf{1.013}$ & $1.895$ & $\mathbf{0.823}$ & $\mathbf{0.022}$ & $\mathbf{0.047}$ & $\mathbf{0.030}$ \\
ResNet18 & - & $1.302
$ & $2.125$ & $1.134$ & $0.0232$ & $0.056$ & $0.038$ \\
UNet & - & $1.938$ & $2.985$ & $1.195$ & $0.028$ & $0.0623$ & $0.038$ \\
\br
\end{tabular}
\end{table*}

\begin{figure*}[!ht] 
\centering
\includegraphics[width=0.9\linewidth]{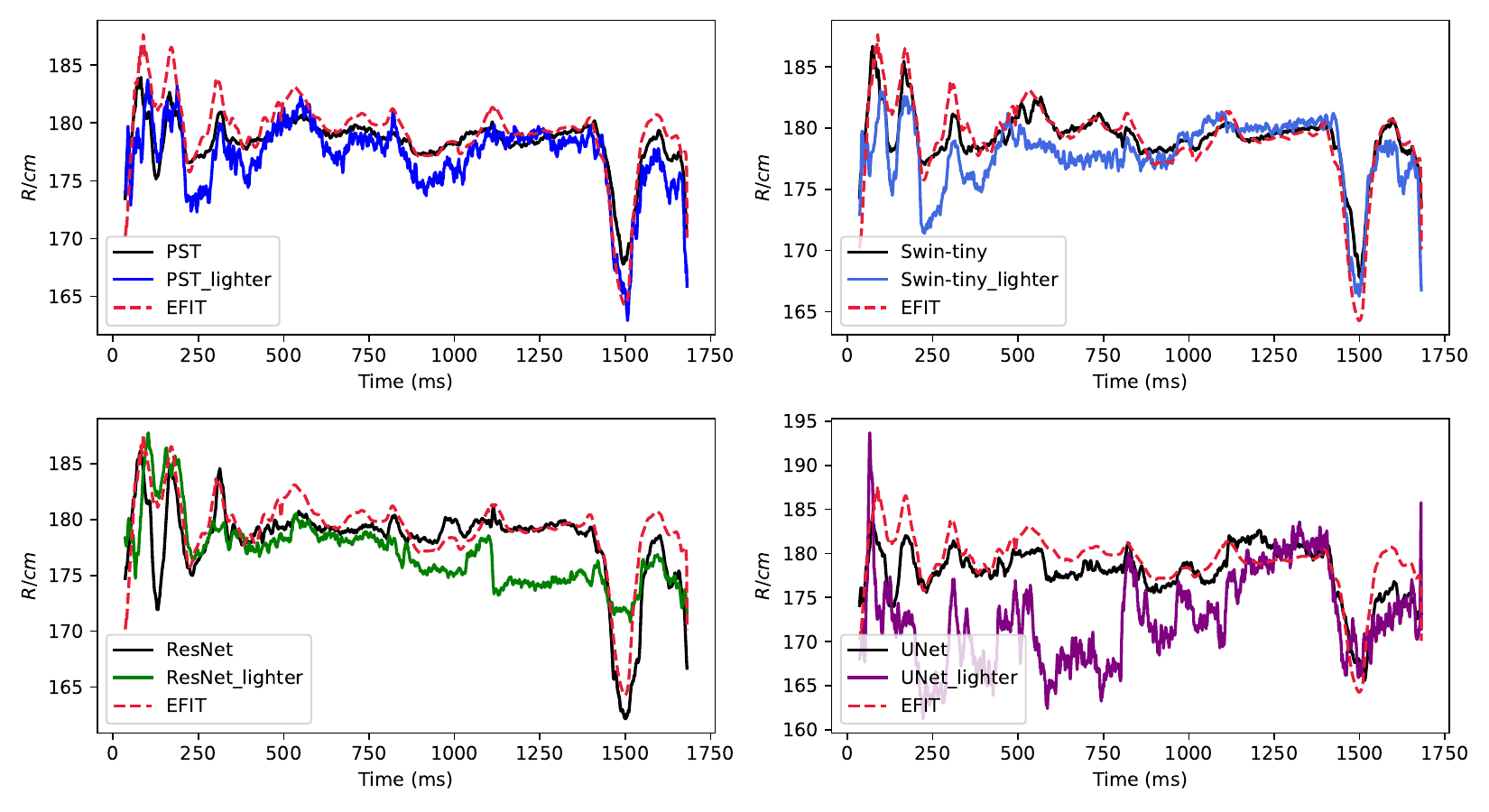}
\caption{Comparison of the horizontal displacement $R$ outputs from different models and the EFIT output after enhancing the brightness and contrast of the overall shot $\#06696$. The black lines represent the inference results of each model on the original image, the red lines represent the EFIT calculation results, and the colored lines are the results after increasing brightness and contrast.}
\label{fig:comparsion of lighter images}
\end{figure*}

Finally, Table \ref{table:offline_comparison} summarizes the comparison between PST-tiny and other models in terms of MAE. It's worth noting that the PST model, upon the incorporation of the Swin Transformer, demonstrates comparable accuracy but significantly faster inference speed over Swin-Tiny. On the contrary, RestNet18 and UNet exhibit reduced accuracy and possibly further deteriorate under conditions of enhanced brightness and high contrastness. Taking the example of shot $\#06696$ in Figure \ref{fig:comparsion of lighter images}, while all models produce comparable results under normal conditions, RestNet18 and UNet turn to perform poorly when the \texttt{convertScaleAbs} function from the OpenCV library is applied with a contrast factor $\alpha$ of 2 and a brightness offset $\beta$ of 10 to adjust the image's contrast and brightness, respectively. 
Instead, the PST model maintains superior stability. Notably, operations such as gas puffing and enhancing the ionization rate often lead to increased density, resulting in over-bright CCD images. Therefore, the robustness of the PST model under challenging conditions underscores its reliability and effectiveness for real-time control.

\subsection{On-Device Results}

\begin{figure}[!thp] 
\centering
\includegraphics[width=0.95\linewidth]{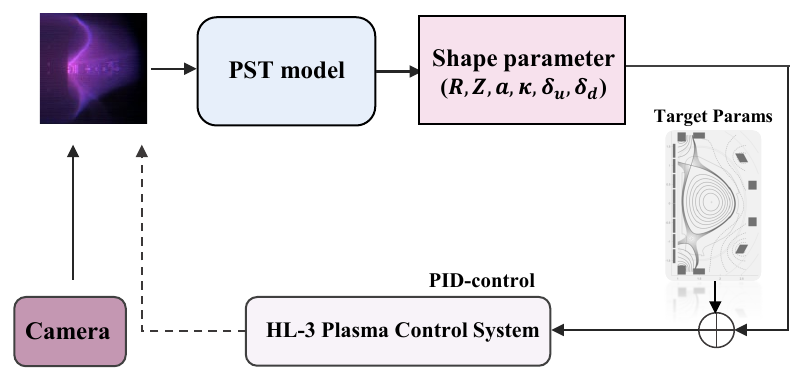}
\caption{The illustration of the PST model serving as a diagnostic means for real-time PID control.}
\label{fig:control diagram}
\end{figure}
\begin{figure}[!tbp]    
    \centering
    \begin{subfigure}[b]{\columnwidth}
        \includegraphics[width=\columnwidth]{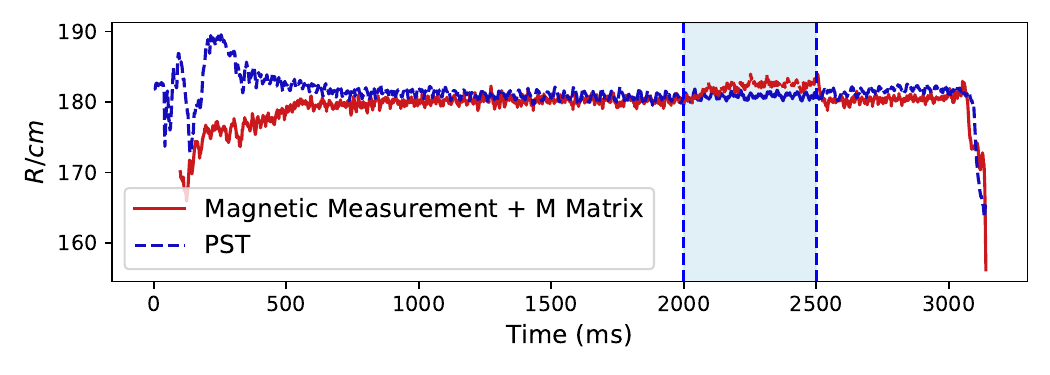}
        \caption{}\label{fig:online_results_a}
    \end{subfigure}
    \hspace{5mm}
    \begin{subfigure}[b]{\columnwidth}
        \includegraphics[width=1\columnwidth]{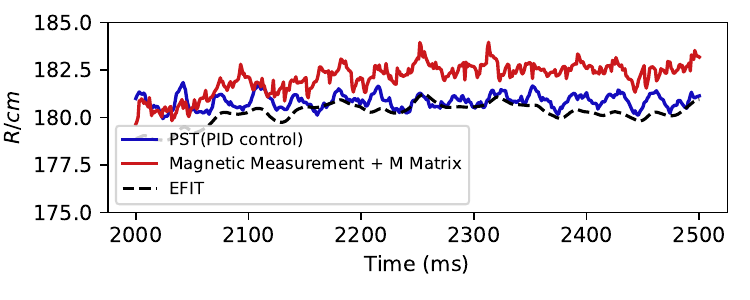}
        \caption{}
        \label{fig:online_results_b}
    \end{subfigure}
\caption{Results of deploying the PST model online and implementing PID feedback control on shot $\#07059$. The blue region represents the phase in which $R$ from the PST is used as input for feedback control. The original input for PID control is obtained using magnetic measurement and the $M$ Matrix method of magnetic measurements. (a) is the real-time detection results of the entire shot, and (b) represents a detailed comparison between the control segment of the PST model and other computational methods.}
\label{fig:online_results} 
\end{figure}
Transitioning from offline simulations to real-time applications, we build a framework as in Figure \ref{fig:control diagram} to seamlessly process captured CCD images and integrate the PST-based outputs into the PCS of HL-3. Notably, the PST model infers from CCD images with camera exposure time and delivers six plasma shape parameters into the PCS through reflective memory. Afterward, the PID controller, a classic feedback control mechanism, reads data from the corresponding memory location every $0.1$ ms and adjusts the control variables according to the evolution of $R$ as well as other plasma shape parameters from magnetic diagnosis, thereby enabling real-time control of plasma shape.

Figure \ref{fig:online_results_a} presents the real-time control result with shot $\#07059$, where, in the duration of $2000$ - $2500$ ms, the horizontal displacement $R$ outputs from the PST model is activated as an input into the PID control system, replacing the magnetic diagnostic $R$. Within the $500$ ms, the PID control maintains a stable control of the plasma. Figure \ref{fig:online_results_b} provides a detailed comparison during this period, incorporating the EFIT calculation method as well, revealing that the PST model's output aligns more closely with EFIT. On the contrary, while the original PID control method is based on magnetic measurements and the $M$ matrix, a systematic deviation (an upward shift of $1$ - $2$ cm) is observed. 
This validates that PST model can effectively utilize the inference results from CCD images as a diagnostic source to control the plasma shape, promising real-time capability with improved accuracy.

\section{Conclusion and Future Research}

In this work, we have developed a lightweight DNN model - PST, for accurately and quickly perceiving CCD plasma images without any manual labeling on HL-3. This model predicts the plasma shape parameters, taking the CCD image as input and outputting six parameters including the radial position $R$ and vertical position $Z$ of the plasma geometric center, minor radius $a$, elongation $\kappa$, upper triangularity $\delta_{u}$, and lower triangularity $\delta_{l}$. 
Specifically, the PST model adapts Poolformer and Swin Transformer towards a lighter-weight design. Meanwhile, we incorporate masked generative distillation, and adopt a multi-task learning framework with the dynamic weight strategy, obtaining high inference accuracy on the PST model. The PST model can predict six parameters within the entire process of $1.8$ ms, with the average MAE for $R$ and $Z$ reaching $1.1$ cm and $1.8$ cm respectively. Furthermore, the PST model manifests the comprehensive adaptability to the visual complexity of the HL-3 device plasma, including overly blurred boundaries, NBI and gas puffing interference, sudden bright spots, and wall hole interference. This specific model can seamlessly integrate with the PCS and effectively support immediate magnetic field control. We have completed $500$ ms PID stable control according to the PST-yielding horizontal displacement $R$ parameter. This preliminary verifies the stability of the PST model in real-time control.

In summary, the results of this research further broaden the potential applications of AI in the field of tokamak plasma control and lay a solid foundation for the development of related technologies in the future.

\section*{Acknowledgments}
This work was in part supported by the National Natural Science Foundation of China under Grant No. U21A20440, the Sichuan Province Innovative Talent Funding Project for Postdoctoral Fellows under Grant No. BX202222, and the Natural Science Foundation of Sichuan Province under Grant No. 2024NSFSC1335.

\section*{Appendix: Details in Swin Transformer}

\begin{figure}[!tp] 
\centering
\includegraphics[width=0.85\linewidth]{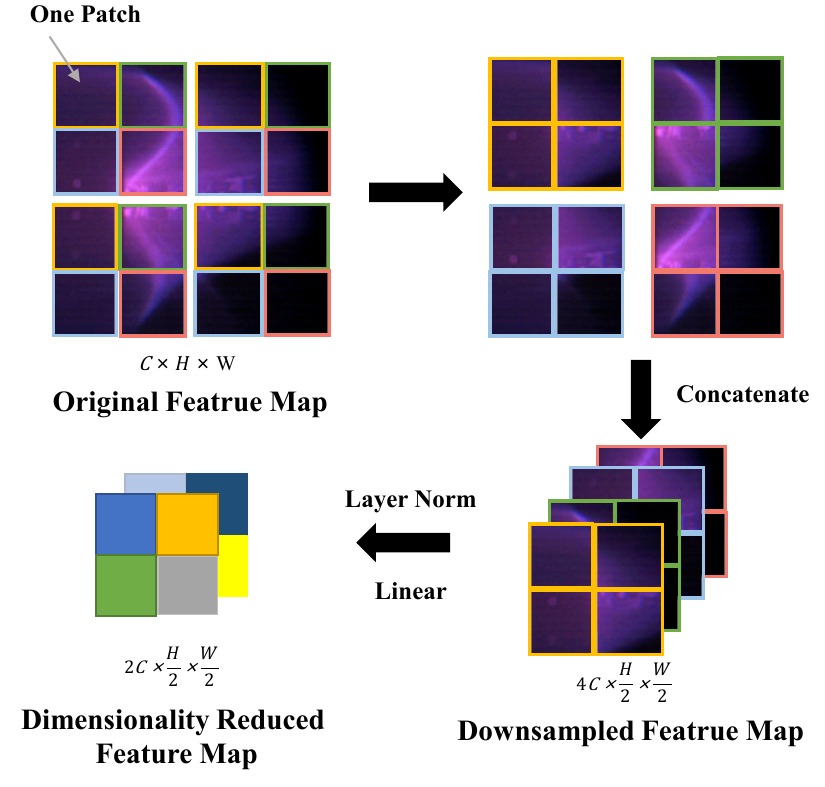}
\caption{Schematic illustration of the Patch Merging process in Swin Transformer.}
\label{fig:patch_merging}
\end{figure}

\subsection*{A. Patch Merging}
Swin Transformer adopts Patch Merging, a downsampling technique to the resolution of the input data while increasing the depth of the information. Within this context, a ``Patch" is defined as the smallest unit in the feature map. To put it differently, a feature map with $14\times14$ pixels comprises $196(=14\times14)$ patches. As illustrated in Figure \ref{fig:patch_merging},  Patch Merging aggregates each cluster of adjacent $n\times n$ patches by depth-concatenation, reducing the dimensions by a factor of $2$. Consequently, the input dimensionality is transformed from $C\times H\times W$ to $4C\times\frac{H}{2}\times\frac{W}{2}$, where $H$, $W$, and $C$ representing height, width, and channel depth respectively. Subsequently, the feature map undergoes a fully connected layer, which adjusts the channel dimension to half of its original size. Through the hierarchical design, it is conducive to capturing more complex features with effectively reduced computational load \cite{liu2021swin}.

\subsection*{B. Self-Attention Block}

\begin{figure}[!tp] 
\centering
\begin{subfigure}{.45\columnwidth}
  \centering
  \includegraphics[width=\linewidth]{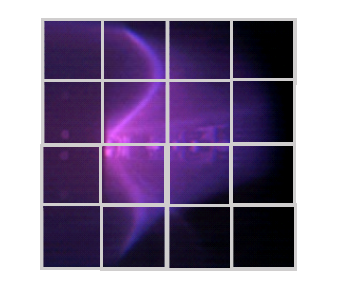}
  \caption{}
  \label{fig:MSA}
\end{subfigure}%
\begin{subfigure}{.5\columnwidth}
  \centering
  \includegraphics[width=\linewidth]{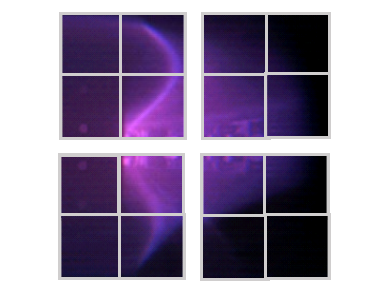}
  \caption{}
  \label{fig:W-MSA}
\end{subfigure}
\newline
\begin{subfigure}{\columnwidth}
  \centering
  \includegraphics[width=.9\linewidth]{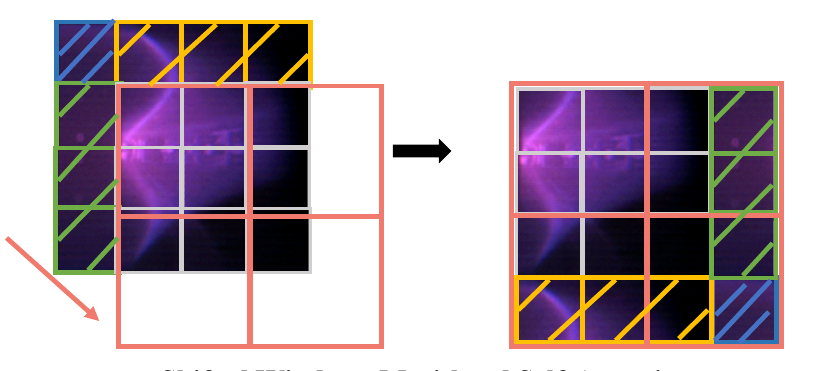}
  \caption{}
  \label{fig:SW-MSA}
\end{subfigure}
\caption{Illustration of Self-Attention Blocks: (a) Multi-head Self-Attention (MSA), (b) Windows-Multi-head Self-Attention (W-MSA), and (c) Shifted Window-MSA (SW-MSA).}
\label{fig:attention_block}
\end{figure}

Traditional Transformers perform attention calculations on a global scale, which results in substantial computational complexity. In contrast, the Swin Transformer confines these calculations within individual windows, thereby achieving a significant reduction in computational load.

As depicted in Figure \ref{fig:attention_block}, compared to the conventional MSA module, the W-MSA divides the feature map into individual windows of size $M\times M$ (with $M = 2$ in this example), and subsequently performs self-attention computations independently within each window. However, the window-limited computations in the W-MSA module inhibit the underlying usefulness of the information transfer between different windows. To circumvent this shortcoming, the Swin Transformer introduces the SW-MSA module, which shifts the window by a factor of $\frac{M}{2}$ both downward and rightward and moves patches into vacant slots to form a complete window for computing efficiency. This innovative strategy effectively resolves the issue of limited information exchange and improves the performance of Swin Transformer \cite{liu2021swin}.

\section*{References}

\bibliographystyle{unsrt}
\bibliography{references} 

\end{document}